\documentclass{aa}  
\usepackage{natbib}
\bibpunct{(}{)}{;}{a}{}{,} 
\usepackage{graphicx}
\usepackage{txfonts}
\newcommand{\kms}{\mbox{${\rm km~s}^{-1}$}}
\newcommand{\htwo}{\mbox{${\rm H}_2$}}

\newcommand{\cc}{{\rm cm}$^{-3}$}

\newcommand{\degree}{$^{\circ}$}
\newcommand{\micron}{$\mu$m}

\newcommand{\msun}{M$_\odot$}

\newcommand{\etal}{\mbox{$et\ al.$~}}

\begin{document}

   \title{The rate and latency of star formation in dense, massive clumps 
in the Milky Way
\footnote{Tables 2,3 and 4 are only available in electronic form
at the CDS via anonymous ftp to cdsarc.u-strasbg.fr (130.79.128.5)
or via http://cdsweb.u-strasbg.fr/cgi-bin/qcat?J/A+A/}
}
   \titlerunning{Rate and Latency of Star Formation}


   \author{M.Heyer  
          \inst{1,2}
          \and
          R. Gutermuth\inst{1}
          \and 
          J.S. Urquhart\inst{2,3} 
          \and
          T. Csengeri\inst{2} 
          \and
          M. Wienen\inst{2} 
          \and
          S. Leurini\inst{2} 
          \and 
          K. Menten\inst{2} 
          \and
          F. Wyrowski\inst{2} }
   \authorrunning{Heyer \etal }
   \institute{Department of Astronomy, University of Massachusetts, 
              Lederle Research Building, Amherst, MA 01003 \\
              \email{heyer@astro.umass.edu}
         \and
             Max Planck Institute for Radio Astronomy, Auf dem H\"ugel 69, 53121 Bonn, Germany
         \and 
             Centre for Astrophysics and Planetary Science, University of Kent, Canterbury, CT2\,7NH, United Kingdom
             }


 
  \abstract 
   {Newborn stars form within the localized, high density regions of molecular clouds.  The sequence and rate at which stars form in dense clumps and 
the dependence on local and global environments are key factors in 
developing descriptions 
of stellar production in galaxies.
   }
   {We seek to observationally 
constrain the rate and latency of star formation in dense massive 
clumps that are distributed throughout the Galaxy and to compare these 
results to proposed prescriptions for stellar production.
   }
   {
A sample of 24\micron-based Class~I protostars 
are linked to 
dust clumps that are embedded within molecular clouds 
selected from the APEX Telescope Large Area Survey of the Galaxy.
We determine the fraction of star-forming 
clumps, $f_*$, that imposes a constraint on the latency of star formation in 
units 
of a clump's lifetime.  Protostellar masses are estimated from models of 
circumstellar environments of young stellar objects from which star formation 
rates are derived.  Physical properties of the clumps are calculated from 870\micron\ 
dust continuum emission and NH$_3$ line emission. 
}
   {Linear correlations are identified between the star formation rate 
surface density, $\Sigma_{SFR}$, and the quantities $\Sigma_{H2}/\tau_{ff}$ 
and $\Sigma_{H2}/\tau_{cross}$, suggesting that star formation is regulated 
at the local scales of molecular clouds. 
The measured fraction of star forming clumps is 23\%.  
Accounting for star formation within clumps that are excluded from our sample due 
to 24\micron\ saturation, this fraction can be as high as 31\%, which is similar to 
previous results.  
Dense, massive 
clumps 
form primarily low mass ($<$ 1-2 \msun) stars with emergent 24\micron\ fluxes 
below our sensitivity limit 
or are incapable of forming any stars for the initial 70\% 
of their lifetimes. 
The low fraction of star forming clumps in the Galactic center relative to those located in the disk of the Milky Way is verified. 
   }
   {}


   \maketitle
%

\section{Introduction}
The formation of young stars and stellar clusters in galaxies follows a sequence of interstellar processes that progressively 
redistributes gas into 
denser, more compact configurations. 
Clouds of primarily molecular gas develop 
from the diffuse, atomic interstellar medium 
in response to 
gravito-magnetic-thermal instabilities within spiral arms or converging flows of warm, neutral material \citep{Kim:2006, 
Heitsch:2006, Dobbs:2014}, or may build-up into larger complexes from the agglomeration of small, pre-existing molecular 
clouds \citep{Dobbs:2008}.
Large molecular clouds typically fragment 
into clumps with enhanced volume and column density that 
comprise 5-10\% of the mass of the cloud 
\citep{Battisti:2014}.
The clumps break down further into localized parcels of gas (cores) 
with even higher volume densities \citep{Beuther:2007, Schneider:2015}.  Single, newborn 
stars emerge from these localized cores 
while young stellar clusters are 
generated if the clump itself is gravitationally unstable \citep{Bate:2003,Vazquez:2009}.
Understanding how each step in this sequence limits the 
rate and yield of newborn stars is a key requirement to the development 
of a complete and predictive description of star formation in galaxies. 
In this study, we address whether star formation 
is regulated by global, galaxy-wide processes, or by the local conditions 
of the gas from which stars condense. 

Several recent studies have addressed this issue using data that spans 
a large range in spatial scales and environmental conditions
\citep{Krumholz:2012, Federrath:2013,Salim:2015}.
\citet{Krumholz:2012} analyzed star formation rates and molecular gas 
properties for molecular clouds in the solar neighborhood, Local group 
galaxies, unresolved disks, 
and starburst galaxies 
in the local and high-redshift 
universe.  With these data, they evaluated several 
proposed star 
formation relationships that describe the variation of star formation rates with 
the amount of gas processed over key timescales, such as the orbital period 
and free-fall time.  The volumetric law, expressed as 
$\Sigma_{SFR}=\epsilon_{ff} \Sigma_{H2}/\tau_{ff}$, 
where 
$\Sigma_{H2}$ is the molecular gas surface density, $\tau_{ff}$ is the free-fall time, and 
$\epsilon_{ff}$ is the star formation efficiency per free-fall time, 
provided the best fit to the full range of data for 
$\epsilon_{ff}$=0.01.  This value for the efficiency is supported by theoretical estimates
\citep{Krumholz:2005, Padoan:2012}. 

\citet{Evans:2014} examined a more recent compilation of young 
stellar objects within the Gould's Belt star forming regions.
They identified a linear relationship between the star formation rate and the amount of dense gas mass that is consistent with a threshold for star formation, 
$\Sigma_{SFR}=f_{dense}\Sigma_{H2}/\tau_{dense}$,
where $f_{dense}$ is the fraction of dense gas within a molecular cloud,
and $\tau_{dense}$ is a characteristic timescale for this dense gas.
Moreover, they found no correlation between the star formation rate and 
mass for cloud extinctions greater than 2 magnitudes and no correspondance 
with the volumetric law. 

In this study, we investigate the star formation rates and gas properties
of 
high density clumps embedded within molecular clouds distributed 
throughout the Milky Way.   
These data extend the number and environment 
of resolved, Galactic star forming regions with which to evaluate the star formation laws. 
The ratio of star forming to quiescent clumps is determined to estimate the time required for a 
typical clump to initiate star formation. 

\section{Data and source selection}
We exploit the 
information within two surveys of the Galactic plane -- the
APEX Telescope Large Area Survey of the Galaxy, (ATLASGAL) \citep{Schuller:2009, Csengeri:2014} 
and the Spitzer MIPSGAL survey 
of 24\micron\ emission \citep{Carey:2009}.  These surveys respectively trace the location and properties of dense clumps and 
color-selected young stellar 
objects in the Milky Way.
ATLASGAL imaged the 870\micron\
thermal dust continuum emission over longitudes 60\degree\ $ > l > $ 280\degree\  
and latitudes $|b| < 1.5$\degree\ with more extended latitude coverage, -2 $< b <$ +1 between 
280\degree\ and 300\degree. The angular resolution of the survey is 19.2\arcsec\ and the surface brightness sensitivity 
is typically 70~mJy/beam.  Two source catalogs have been constructed from the ATLASGAL image data. 
\citet{Csengeri:2014} tabulated a list of compact objects based on the Gaussclump decomposition algorithm 
which samples localized emission peaks \citep{Stutzki:1990}. 
\citet{Urquhart:2014a} applied the  
the SExtractor algorithm to define spatially connected structures \citep{Bertin:1996}.  Such objects are 
generally more extended and circumscribe the localized emission peaks.  In this study, we 
use the \citet{Urquhart:2014a} catalog, which includes a
set of mask images 
that define the set of pixels tagged to each ATLASGAL source. 

The MIPSGAL Survey is a Legacy Program of the Spitzer Space Telescope that imaged the 24 and 70\micron\ emission along the Galactic 
plane over the areas -68\degree\ $< l <$ 69\degree, $|b| <$ 1\degree and -8\degree $< l <$ 9\degree, $< |b| <$ 3\degree\ to a sensitivity depth of 1~mJy
\citep{Carey:2009}.  \citet{Gutermuth:2015} constructed a 24\micron\ point source catalog from the MIPSGAL images as well as a lower signal to noise
archive set of point sources.  
Here, we restrict our analysis to the more reliable set of catalog sources.  
Owing to limited sensitivity in the MIPS detectors, the catalog does not include 70\micron\ photometry. 
An auxilary data product of the MIPSGAL source catalog used in this study is the set of 
differential completeness decay data cubes for each image tile that tabulate the 
point source detection completeness within 1 arcmin$^2$ pixels 
as functions of flux and position given the local background \citep{Gutermuth:2015}. 

\subsection{Selecting ATLASGAL sources}
We examine dust sources (hereafter, clumps) from ATLASGAL within the Galactic longitudes 
60\degree\ $> l >$ 300\degree\
and 
latitudes, $|b|<$ 1\degree. 
There are 9494 ATLASGAL sources located within these boundaries.

The effects of saturation and a spatially varying background within the MIPSGAL images 
must be considered when 
evaluating star formation activity within a clump.  
Depending on the local background level, the Mips 24\micron\ detectors saturate for points sources with flux $\sim>$2~Jy. 
The MIPSGAL source catalog constructed by \citet{Gutermuth:2015} excludes 
such positions from analysis 
and flags these locations within the differential completeness decay data cubes.  
Any ATLASGAL clump whose solid angle 
subtends any flagged, saturated MIPSGAL pixel is excluded from our analysis.  
Of the original sample of 9494 ATLASGAL clumps, 
1487 overlap with 
saturated MIPSGAL pixels and are removed from the clump list. 

\begin{figure}
\centering
\includegraphics[width=\hsize]{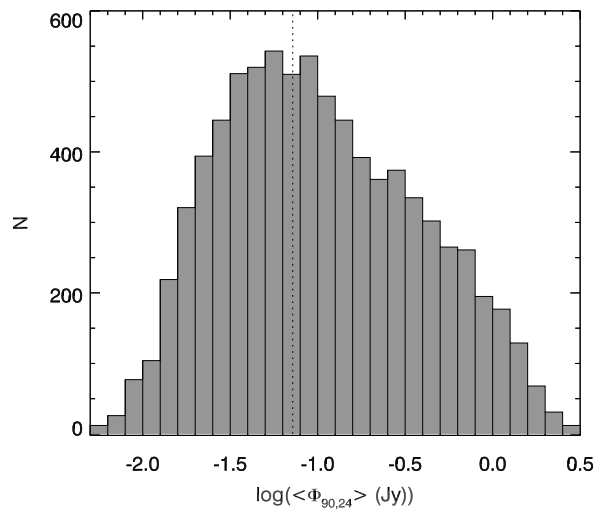}
\caption{Distribution of $<\Phi_{24,90}>$ values for ATLASGAL clumps with $|b| <$ 1\degree\ and containing no 
saturated pixels at 24\micron.  
The vertical dotted line marks the maximum value of $<\Phi_{24,90}>$ that defines the clump sample
and reflects the flux at which the MIPSGAL source catalog is at least 90\% complete in these areas. 
}
\label{d90}
\end{figure}

The second consideration in our selection of ATLASGAL clumps is source completeness. 
The 24\micron\ background measured by MIPSGAL exhibits large spatial variations 
throughout the Galactic plane.  This variation arises from 
extended nebulosity excited by ultraviolet radiation that heats interstellar dust grains, which cool by reradiating 
this energy in the mid and far infrared wavelength bands.  The effect of this background is to 
reduce the local sensitivity to faint 24\micron\ point sources.  As ATLASGAL clumps can be embedded within such nebulosity, 
the corresponding sensitivity to associated YSOs within the projected clump area is variable across the Galactic plane. 
To characterize the 24\micron\ background within an ATLASGAL clump and the sensitivity to point sources within its 
area, we define 
\begin{equation}
<\phi_{24,90}>=\frac{\int \phi_{24,90} d\Omega}{\int d\Omega}, 
\end{equation}
where $\phi_{24,90}$ is the interpolated value from the differential completeness decay cubes to 90\% and 
the integral is over the solid angle of the ATLASGAL source.
Figure~\ref{d90} shows the distribution 
of $<\phi_{24,90}>$ for 8007 ATLASGAL clumps without 24\micron\ saturation.
To establish a base sensitivity to YSOs for all ATLASGAL clumps, a source completeness requirement is imposed such that 
$<\phi_{24,90}>$ is less than 72~mJy ([24]=5). 
This limit enables our analysis to be 
at least 90\% complete to protostars with mass greater than 2~\msun\ at a distance of 3.9~kpc 
(see Section~4).
 A total of 3494 ATLASGAL sources satisfy this requirement 
and comprise our final sample of clumps.

\subsection{Selecting young stellar objects from MIPSGAL}
The \citet{Gutermuth:2015} MIPSGAL source catalog is a compilation of point sources extracted from 24\micron\ image product.  In addition to the positional centroids, 24\micron\ photometry, and quality flags,  the table 
also includes photometry from the nearest entry in the 2MASS, GLIMPSE, and WISE catalogs if the source positions from  these 
catalogs fall within 2\arcsec\ of the MIPSGAL positional centroid. 
The availability of source fluxes from the near to mid-infrared bands allows one to identify candidate young stellar objects based on the 
shape of the spectral energy distribution between 1-25\micron. 
In this study, we are interested in the early 
stage YSOs that should still be located within the clumps from which they formed.  
Such early stage YSOs exhibit rising spectral energy distributions 
through the mid-IR bands 
(Class~I)
\citep{Lada:1992}.

Using the infrared color criteria for Class~I or 
deeply embedded protostars 
described by \citet{Gutermuth:2009}, 13406 candidate early-stage YSOs 
are selected from the MIPSGAL catalog with $|l|< 60$ and $|b|<1$.  
A fraction of these candidate YSO sources could be AGB stars, which exhibit similar colors as early stage YSOs.  
\citet{Robitaille:2008} estimated 
30-50\% of the GLIMPSE-based red sources are AGB stars and proposed a magnitude and color threshold
to distinguish between AGB stars and YSOs. 
 In this study, we do not directly apply this 
color-magnitude cut since we are interested in the angular coincidence between 
ATLASGAL clumps and YSOs.  It is 
possible, yet unlikely, that AGB stars are aligned with the small area subtended by such clumps.   
Nevertheless, an examination of the photometry 
for the selected sources satisfies these magnitude and color requirements. 

\subsection{Selection biases}
To address the biases generated by the respective selection requirements for ATLASGAL clumps and MIPSGAL YSOs, 
we
compare our ATLASGAL sample and YSO populations to those used in previous studies.  \citet{Urquhart:2014b} 
compiled a complete sample of massive star forming clumps in the Galaxy by matching sources within 
the ATLASGAL 
catalog to various tracers of massive star formation.  These tracers include the Red MSX Source (RMS) Survey \citep{Lumsden:2013}, 
the Methonal Multibeam (MMB) Survey \citep{Green:2009}, and compact HII regions extracted 
from the Cornish Survey \citep{Purcell:2013}.  
Cross linking our sample of 3494 ATLASGAL clumps with those associated with these signatures of massive 
star formation,
we find the following matched fractions: 23/329 (7\%) to RMS, 63/529 (12\%) to MMB, and 
21/556 (4\%) to compact HII regions.   The low overlap fraction arises from both saturation within the 
MIPSGAL images and bright backgrounds in the vicinity of massive stars.  Specifically, 43\%, 42\%, and 58\% of 
the ATLASGAL sources examined by \citet{Urquhart:2014b} are discarded in our study 
 due to saturated pixels at 24\micron\ for RMS, MMB, and compact HII regions 
respectively.  Similarly, 50\%, 46\%, and 38\% are removed from our sample of clumps owing to 
the source completeness requirement. 
The requirements imposed on our sample of ATLASGAL clumps introduces a bias that mostly excludes sites
of massive star formation 
as traced by the RMS, MMB, and compact HII region CORNISH catalogs. 

\citet{Robitaille:2008} constructed a list of 18337 red infrared objects based on the color selection [4.5]-[8.0] $>$ 1
for sources with $\sim$6:1 signal to noise in these bands.  These red objects are comprised of unresolved planetary nebulae, 
AGB stars, YSOs, and background galaxies.  Using targeted MIPSGAL 24\micron\ photometry on these sources, they further refined 
their sample of YSOs based on [4.5]$>$ 7.8 and [8.0]-[24] $\ge$ 2.5. From these cuts, they estimate a total of 11473 YSOs.  
Within this sample, 2184 do not have 24\micron\ photometry owing to bright background or local saturation of the 
Mips detectors. 

The candidate YSOs derived 
from the MIPSGAL catalog are matched to  2744 (23\%) of the Glimpse-based YSOs.  If we impose the same IRAC 
selection criteria used by \citet{Robitaille:2008},
only 4318 of the MIPSGAL-based YSOs would have been selected by \citet{Robitaille:2008}.  From this subset, 2739 (63\%) match the YSO list of \citet{Robitaille:2008}.  
However, the YSOs identified by \citet{Robitaille:2008} include both early and intermediate stage systems. 
We have classified the \citet{Robitaille:2008} sample using the \citet{Gutermuth:2009} criteria 
assuming 10:1 signal to noise ratio in the MIPSGAL fluxes (not tabulated in the \citet{Robitaille:2008} table).
This classification results in 3887 Class~I, 39 deeply embedded, and 
4193 Class 2 objects.  Of the 3926 Class~I and deeply embedded set, 59\% are matched to the YSO sample in 
this study. 
When analyzed self-consistently, 
there is $\sim$60\% overlap between the Class~I/embedded MIPSGAL-based sample of YSOs and those 
of \citet{Robitaille:2008}. 
A large number (11091) of early stage MIPSGAL YSOs 
are not matched to the GLIMPSE-based YSO sample. 
This mismatch suggests that the YSOs keyed 
on 24\micron\ detection represent a different population of YSOs than those defined 
by \citet{Robitaille:2008}.  This population is likely to be more deeply embedded within the dense clump from which it 
condensed and therefore, represents an earlier stage of YSO evolution. 

\section{Linking young stellar objects to ATLASGAL clumps}
\begin{figure*}
\centering
\includegraphics[width=\hsize]{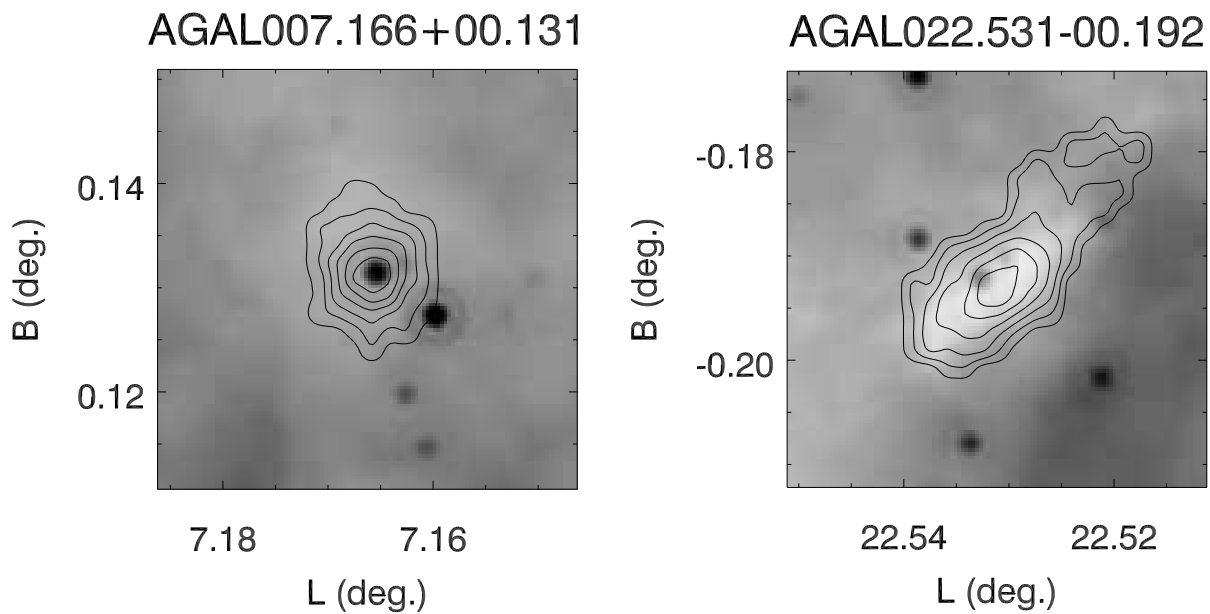}
\caption{
(left) Star forming ATLASGAL clump, AGAL007.166+00.131, associated with a MIPSGAL selected YSO.  The background 
image is the MIPSGAL 24\micron\ image, the contours show the masked 870\micron\ intensity.
The YSO is located at the peak of the dust emission. 
(right) Quiescent clump, AGAL022.531-00.192.  The faint 24\micron\ source projected onto the 
dust emission is not identified as a YSO. 
}
\label{link_examples}
\end{figure*}

To evaluate the star formation activity within this sample of clumps, it is 
necessary to associate early-stage YSOs to ATLASGAL sources. 
A candidate YSO is linked to an ATLASGAL clump if its l,b coordinate lies within the solid angle of the clump as 
defined by its SExtractor mask.  A YSO located just outside of the mask area is not considered to be resident 
within the clump but may represent a later stage of YSO evolution. 
More than one YSO may be linked to a single clump.   Examples of a star-forming clump and a non star-forming, 
quiescent clump, in which no YSO is linked, are 
shown in Figure~\ref{link_examples}.   In many cases, a MIPSGAL point source is coincident with an ATLASGAL clump
but its infrared colors are consistent with a main-sequence star.  

The fraction of star-forming clumps, $f_*=N_*/N_T$, is 0.23, where 
$N_*$ is the number of ATLASGAL clumps with at least one early-stage YSO and $N_T$=3494, is 
the total number clumps in the sample.
This fraction does not significantly change if the completeness 
limit requirement is relaxed such that all 8007 ATLASGAL clumps within the searched 
area with no saturated pixels are examined.  In this case, 22\% of the
 ATLASGAL clumps are associated with a 
MIPSGAL YSO.  While more star forming clumps are identified within this larger sample, the added ATLASGAL 
sources with brighter backgrounds (higher values of $<\Phi_{24,90}>$) preclude 
detections of fainter YSOs.  To maintain a well-defined sample, we only consider the 3494 
clumps with the imposed background flux limit. 

The offset, ${\Delta}r$, between the YSO position and the position of 
peak 870\micron\ 
intensity is calculated for each star forming clump.  This offset is 
normalized by the clump radius.  The distribution of normalized 
offsets is strongly peaked between 0 and 0.1 and
affirms the early-stage status
of YSOs linked to these clumps. 

Several previous studies have examined the fraction of star forming ATLASGAL clumps using other star forming tracers. 
\citet{Contreras:2013} used mid infrared, color-selected data from the MSX mission and found a fraction of star forming clumps of 
40\% over the longitude range 330 $< l <$ 21.  \citet{Csengeri:2014} compiled a list of compact ATLASGAL 
sources and linked both WISE and MSX based protostars based on angular proximity to the dust peak. 
They found 30\% of the compact clumps were linked to YSOs identified within the WISE source catalog with colors 
similar to those used in this study and 33\% of star-forming clumps when including MSX sources in regions where WISE 
saturates.

\begin{figure*}
\centering
\includegraphics[width=\hsize]{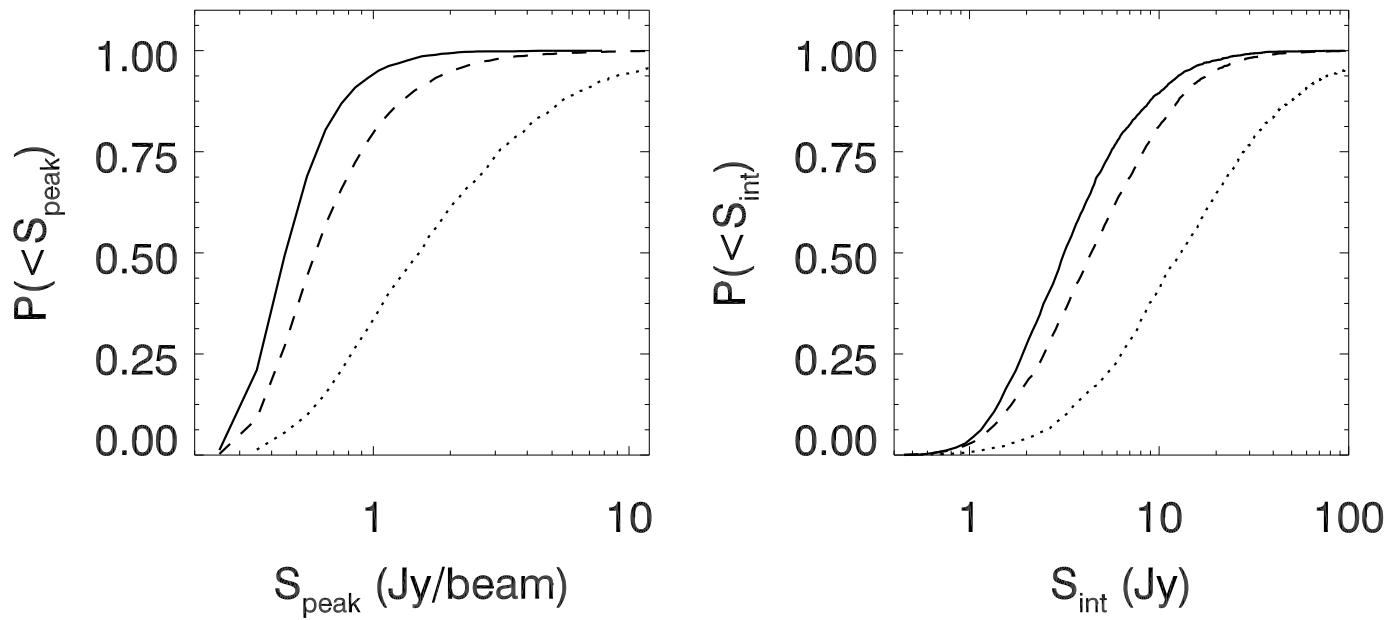}
\caption{
(left) Cumulative distributions of peak flux intensity of ATLASGAL clumps
with MIPSGAL saturation (dotted line), with $<\Phi_{24,90}>$ greater than 72~mJy (dashed line)
and with $<\Phi_{24,90}>$ less than 72~mJy (solid black line).  
(right) Cumulative distributions of integrated flux density of ATLASGAL clumps for these same 
groups. 
}
\label{atlasgal_sources_select}
\end{figure*}

The discrepancy between the fraction of star-forming clumps found in these 
earlier studies and this MIPSGAL 
based study, lies primarily in the sample of 1487 ATLASGAL clumps discarded due to saturation 
in the 24\micron\ band.  Figure~\ref{atlasgal_sources_select}
shows the cumulative distributions of peak and integrated flux densities 
for ATLASGAL sources with 24\micron\ saturation (dotted line), 
with bright backgrounds $<\Phi_{24,90}>$ greater than 0.072~Jy (dashed line), 
and 
with $<\Phi_{24,90}>$ less than 0.072~Jy (solid line).
Clumps with 24\micron\ saturation have larger 870\micron\ peak flux densities 
than the other groups.  \citet{Csengeri:2014} demonstrate that clumps with large peak flux densities have a higher 
probability of being associated with star formation activity.  The number of 
clumps discarded due to saturation that may be linked to YSOs, $N_{*,c}$, is estimated by the integral 
\begin{equation}
 N_{*,c}=\int P(S_{peak})N(S_{peak})dS_{peak} = 755, 
\end{equation}
where $P(S_{peak})$ is the probability of a star-forming core within bins of the peak flux density (see Figure~18 in 
\citet{Csengeri:2014}). 
A corrected value for $f_{*,c}$ that accounts for star formation in these saturated zones is 
$f_{*,c}=\frac{(791+755)}{(3494+1487)}$ = 0.31. This corrected value is more aligned 
with earlier estimates by 
\citet{Contreras:2013} and \citet{Csengeri:2014}. 

\begin{figure*}
\centering
\includegraphics[width=\hsize]{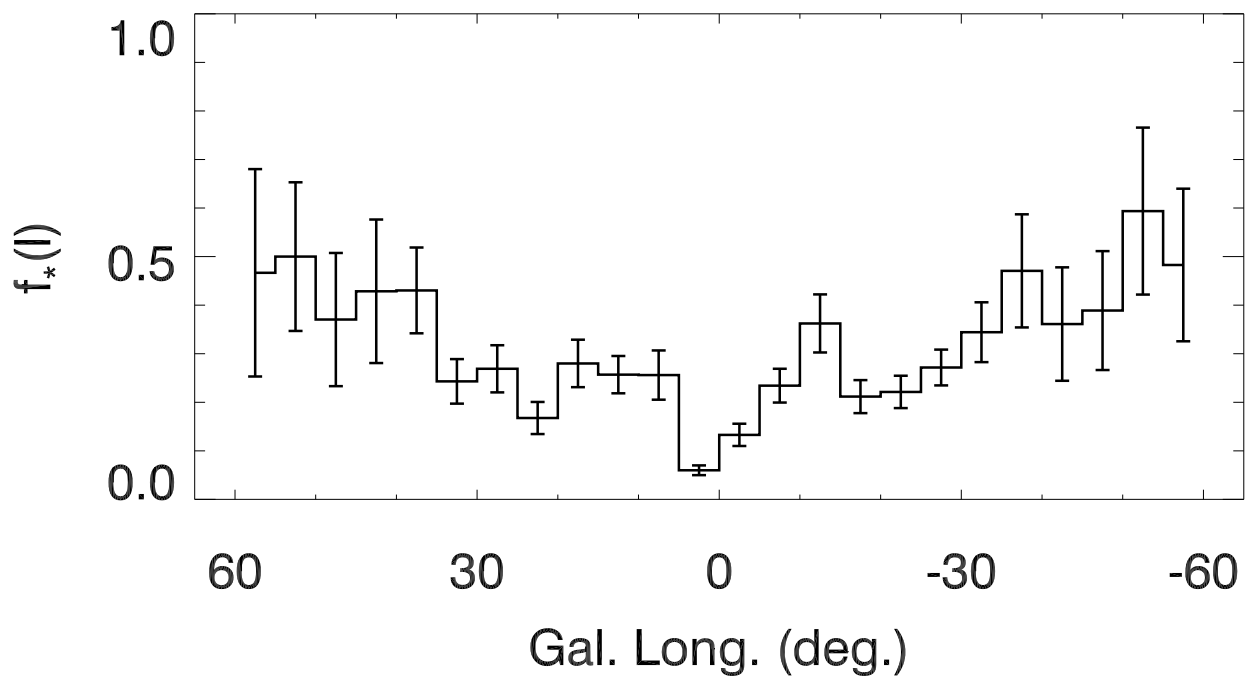}
\caption{
Fraction of star forming clumps, $f_*$,  
with Galactic longitude.  The Central Molecular Zone ($|l|<$2.5) has a significantly smaller fraction of star forming 
clumps than in the Galactic disk. 
}
\label{nyso_gl}
\end{figure*}

The derived fraction of star forming clumps is a global value over the examined area of the surveys.
However, $f_*$ may vary with position in the Galaxy and local environmental conditions.  
Figure~\ref{nyso_gl} shows 
the variation of $f_*$ with Galactic longitude. 
The displayed error-bars reflect the propagation of counting statistics in each bin.  The apparent rise of 
$f_*$ towards both longitude 
ends of the survey results from the limited number of clumps in these segments and therefore, is not 
statistically significant. 
However, the minimum of $f_*$ in the bin centered at $l$=2.5\degree\ is 
significantly below the other bins.
Such reduced star formation 
activity in this region has been previously recognized \citep{Guesten:1983,
Longmore:2013, Urquhart:2013, 
Csengeri:2014}.   \citet{Kruijssen:2014} have examined proposed star formation suppressants
that could be active within
the Central Molecular Zone (CMZ).  The primary limitation appears to be a high volume density 
threshold for star formation ($>$ 10$^7$ \cc) set by the elevated turbulent velocity dispersions in the CMZ \citep{Krumholz:2005}.  
This threshold is much larger than the mean volume densities of clumps (discussed in \S5) located within or near the CMZ. 

\begin{figure*}
\centering
\includegraphics[width=\hsize]{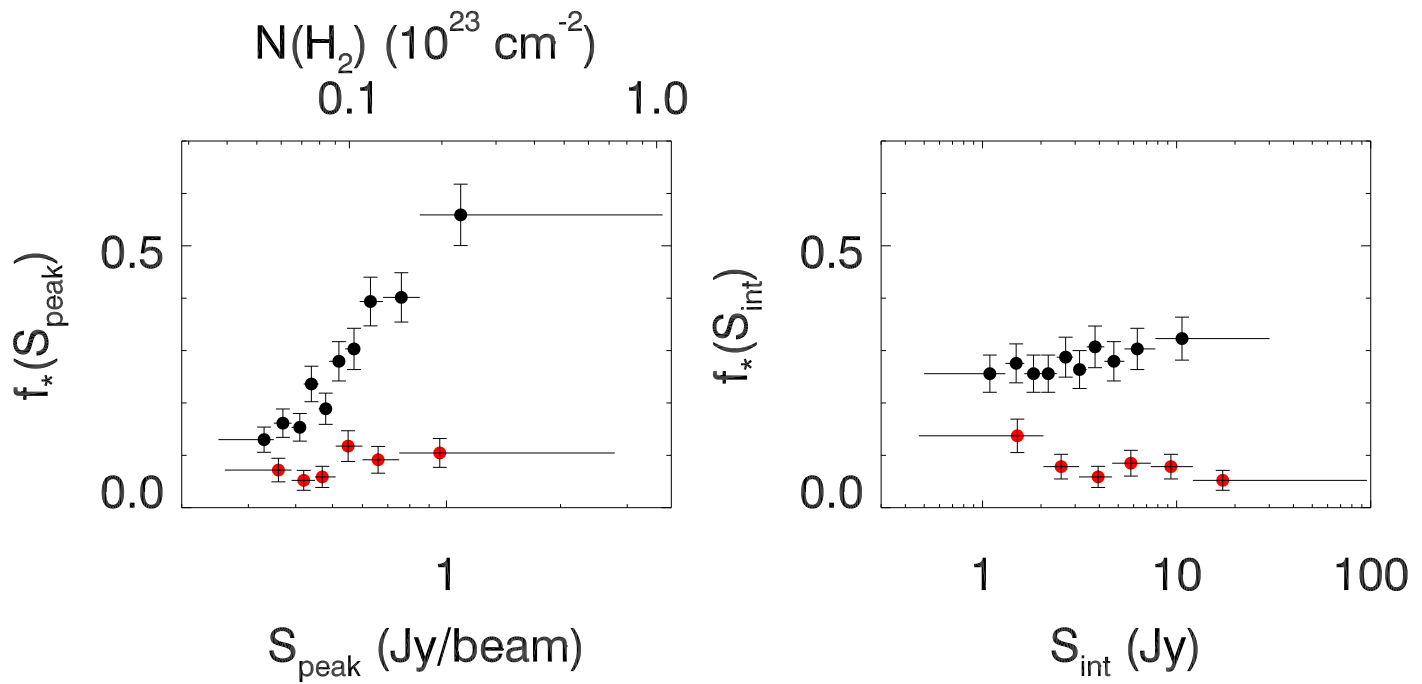}
\caption{(left) Variation of $f_*$ with peak flux density and (right) integrated flux
for ATLASGAL clumps 
within the disk (black) and Galactic center (red).  The bin sizes are adapted to include equal 
number of clumps. The horizontal error bars show the bin widths.  The vertical error 
bars reflect the dispersion of values about the mean of the sample. 
The top x-axis of the left side plot shows the corresponding gas column 
density assuming a dust temperature of 20~K. 
}
\label{nyso_ncolC}
\end{figure*}

The local environment of a given clump is expected to have a role in regulating star formation. 
In particular, regions with higher volume and column density should be more susceptible to 
forming newborn stars as self-gravity begins to dominate over magnetic and turbulent pressure.  Figure~\ref{nyso_ncolC} 
show the variation of $f_*$
 with 870\micron\ peak flux density and integrated flux.  Clumps within the disk and 
Galactic center ($|l|<$ 5\degree) are shown separately.  
The corresponding values of \htwo\ column density 
are shown in the top axis assuming a dust temperature of 20~K and dust opacity of 1.85 $cm^2/gm$ at 
870\micron\ 
\citep{Ossenkopf:1994} and a gas to dust ratio of 100.  
Clumps located in the disk show an increasing 
frequency of star formation with peak flux density.  
Curiously, no such rise is measured for clumps in the Galactic center region.
The fraction of star forming clumps is constant with integrated flux for both disk and Galactic center clumps 

\section{MIPSGAL sensitivity to protostellar mass}
\subsection{Comparison to \citet{Robitaille:2006} models}
To obtain a better understanding of the candidate protostellar objects identified by MIPSGAL,
we have examined the radiative transfer models of YSO environments (central star, disk, envelope) 
and emergent spectral energy distributions computed 
by \citet{Robitaille:2006}. The set of 20,700 models span a range of stellar mass, radius, and surface temperature,
accretion rates from the disk and envelope, 
and different stages of protostellar evolution.  
For each model, spectral energy distributions are calculated for 10 inclination angles relative to the disk axis. 
The model SEDs 
are convolved to commonly used filter response functions to generate broadband fluxes within 
50 different sized apertures for each inclination angle at a fixed distance of 1~kpc. In our analysis, 
we only consider model photometry in the largest aperture of 100,000 AU.  Using somewhat smaller apertures ($>$20,000 AU)
provided by \citet{Robitaille:2006} did not affect our results. 

The YSO model broad band fluxes are used to 
determine early-stage Class~I, Class~II, and late-stage, Class~III objects as applied to the MIPSGAL 
data in \S2.2.  \citet{Robitaille:2006, Robitaille:2008} 
define 3 stages of YSO evolution based on the envelope and disk accretion rates relative to the stellar mass:
Stage~1 has $\dot{M}_{env}/M_* > 10^{-6}$ yr$^{-1}$; 
Stage~2 has $\dot{M}_{env}/M_* < 10^{-6}$ yr$^{-1}$ and ${M}_{disk}/M_* > 10^{-6}$;
Stage~3 has $\dot{M}_{env}/M_* < 10^{-6}$ yr$^{-1}$ and ${M}_{disk}/M_* < 10^{-6}$. 
Table~\ref{table1} shows the number of models in each YSO Stage that are classified as 
Class~I, Class~II, and Class~III spectral energy distributions.
Model YSOs that are classified as Class~I objects are 
primarily linked to Stage~1 YSO models.  
However, a significant number of Stage~1 model YSOs 
are classified as Class~2 objects (37\%).   These Stage~1 models exhibit blue infrared colors and follow 
the locus of Stage 2 sources within a plot of IR spectral index and disk accretion date illustrated 
in Figure~11 of \citet{Robitaille:2006}. 
\begin{table}
\caption{Protostellar stage vs color-based classification of \citet{Robitaille:2006} models
}     
\label{table1}      
\centering         
\begin{tabular}{c c c c}  
\hline\hline              
 &  Stage 1 & Stage 2 & Stage 3 \\   
\hline                        
Class I          & 54253  & 8055  & 639 \\   
Class II         & 33940  & 52225 & 7804 \\
Class III        &   72 & 3399 &  12784 \\
\hline                      
\end{tabular}
\end{table}

\begin{figure*}
\centering
\includegraphics[width=\hsize]{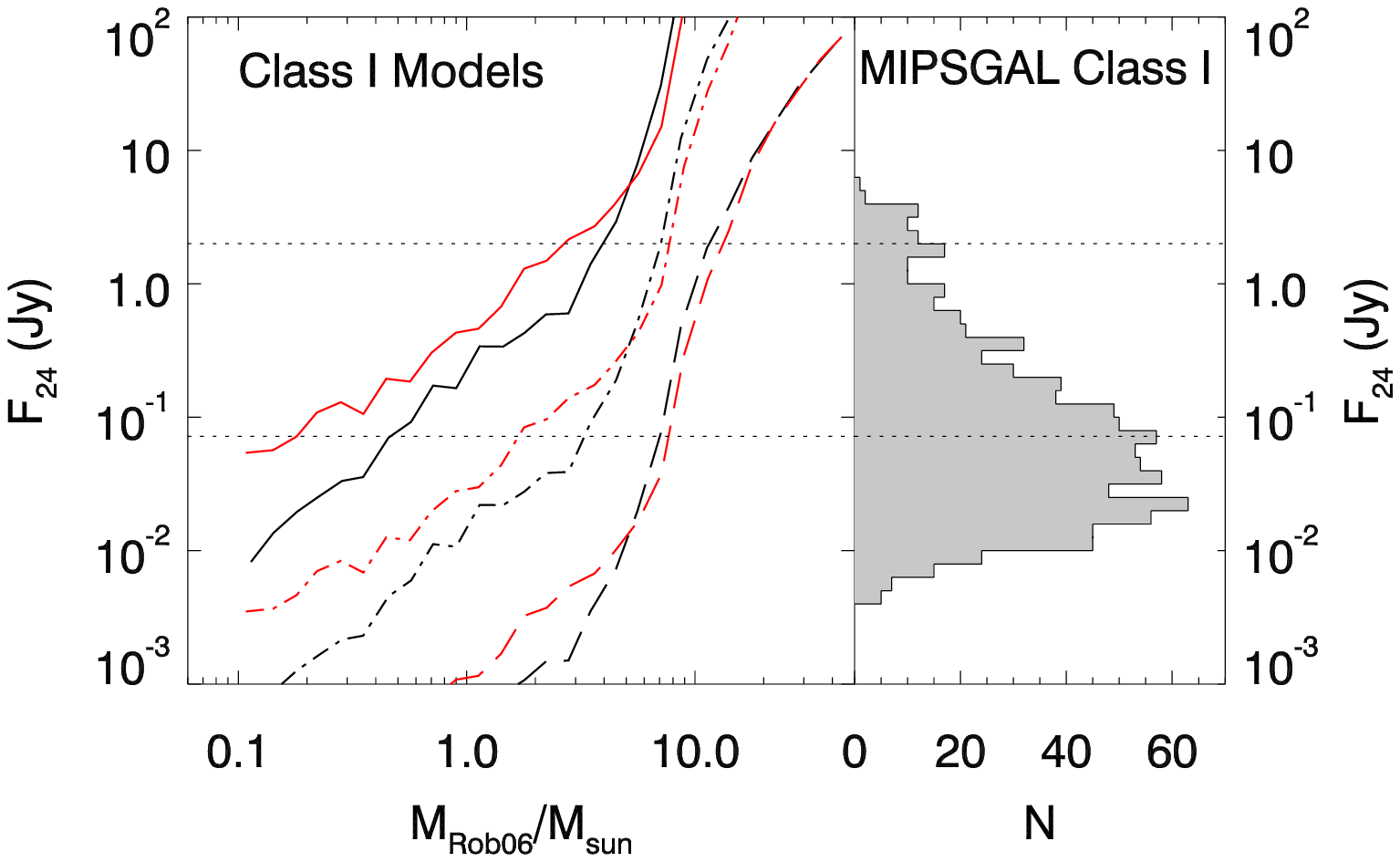}
\caption{(left) Variation of mean 24\micron\ flux within bins of  
protostellar mass and mass accretion rates $10^{-9} \le \dot M_{disk} < 10^{-7} M_\odot yr^{-1}$ (black) and 
$10^{-7} \le \dot M_{disk} < 10^{-5} M_\odot yr^{-1}$ (red)
 derived from the YSO 
models of \citet{Robitaille:2006} and 3 distances (1~kpc (solid line), 
4~kpc (dot-dashed lines) and 20~kpc (dashed lines).  (right) Distribution 
of 24\micron\ fluxes for early stage YSOs linked to ATLASGAL clumps. 
The upper horizontal
dotted line corresponds to the flux level at which the MIPS detectors saturate.
The lower horizontal line shows the flux at which MIPSGAL 
is at least 90\% complete for this sample of ATLASGAL clumps.
}
\label{f24_yso}
\end{figure*}

The sensitivity of our YSO sample to stellar mass can be partially 
constrained by examining the 
distribution of 24\micron\ flux density values.  Figure~\ref{f24_yso} shows the run 
of mean 24\micron\ flux with stellar mass for models classified as Class I objects 
for distances 1, 4.0, and 20 kpc and 
2 bins of disk accretion rates: 
$10^{-9} < \dot M_{disk} < 10^{-7}$ $M_\odot yr^{-1}$ and 
$10^{-7} < \dot M_{disk} < 10^{-5}$ $M_\odot yr^{-1}$.
Attenuation of the 24\micron\ flux by interstellar dust grains distributed over 
these distances is not included. 
Also shown is the distribution of measured 24\micron\ fluxes for the full sample 
of Class~I, MIPSGAL YSOs.  The models show that fluxes from protostars with masses greater 
than 12~\msun\ exceed the 2~Jy saturation limit of the Mips detectors at a distance 
of 20~kpc.  Since this distance is likely an upper limit to any protostar, 
we assign 12~\msun\ as an upper protostellar mass limit to which the MIPSGAL catalog is sensitive. 
The plot also shows that the selected YSO sample is at least 90\% 
complete to protostellar masses of 0.2-0.4~\msun\ at a 
distance of 1~kpc, 1.5-3~\msun\ at 4~kpc, and 7~\msun\ at 20~kpc. 

\begin{figure*}
\centering
\includegraphics[width=\hsize]{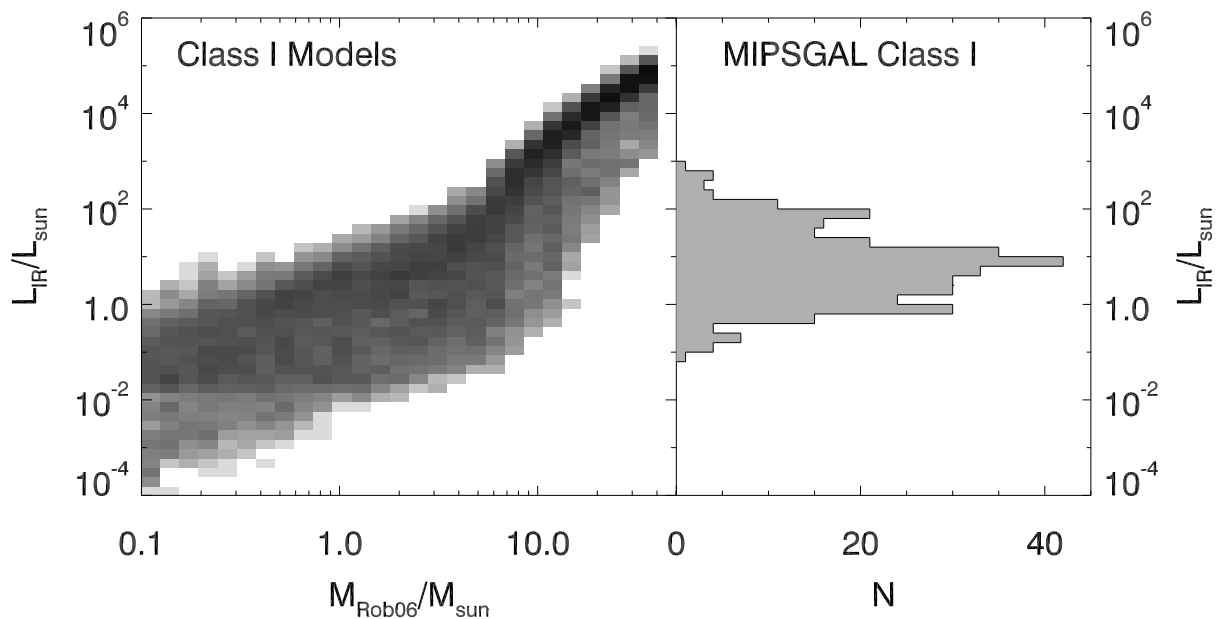}
\caption{(left) 2D histogram, $\zeta(M,L_{IR})$, of mass and infrared luminosity 
for the \citet{Robitaille:2006} YSO 
models with  -8 $< log(\dot{M}_{disk}) <$ -5, integrated over all incidence angles 
 and classified as Class~I/embedded objects.  The color scale is logarithmic and ranges from
0 (white) to 1000 (black) models per bin. (right) Distribution of $L_{IR}$ for 290 Class~I, MIPSGAL YSOs linked 
to 219 ATLASGAL clumps in our sample.
}
\label{lbol_yso_mc}
\end{figure*}

The YSO luminosity can also be used to constrain 
YSO masses.  
The emergent bolometric luminosity from a protostellar 
environment depends on many factors including the disk accretion rate and mass of the 
central, developing star.  
Consequently, the protostellar luminosity 
exhibits a large range of values for a given central star mass.  A significant 
fraction of the protostellar 
bolometric luminosity is emitted at wavelengths longer than 50\micron. 
Since the YSO photometry from the MIPSGAL catalog 
is limited to wavelengths less than 25\micron, we restrict our analysis to the mid-infrared luminosity, $L_{IR}$, 
between 2 and 24\micron\ from the available 2MASS, GLIMPSE, and MIPSGAL photometry. 
Figure~\ref{lbol_yso_mc} shows the variation of 
$L_{IR}$, with 
central, stellar mass, $M_{Rob06}$, 
as a 2D-histogram, $\zeta(M,L_{IR})$, for all models identified as Class~I objects with model disk accretion rates 
in the range,
$10^{-8} < \dot{M}_{disk} < 10^{-5}$ $M_\odot yr^{-1}$ and integrated over all incidence angles.
The adopted range of $\dot{M}_{disk}$ is taken from recent observational 
studies of pre main sequence 
evolution that derive a comparable range of mass accretion rates for Class~I objects \citep{Garatti:2012, Antoniucci:2014}. 
For a given central, stellar mass, there is a large spread of $L_{IR}$ values  
reflecting the varying 
disk accretion rates and range of model time-steps.  

Distances are available for 510 of the ATLASGAL clumps from 
the study by \citet{Wienen:2015} or cross-linking ATLASGAL sources with Bolocam Galactic plane dust sources with distances 
\citep{Ellsworth-Bowers:2015}.  For the \citet{Wienen:2015} sample, the measured velocity and positions 
and near/far side assignment for a given clump 
were used to derive a distance using the \citet{Reid:2009} rotation curve.
Of this sample, there are 219 star forming clumps which are associated with 
290 Class~I YSOs.   The median distance for all clumps in our sample is 3.9~kpc. The most distant 
clump is 25.2~kpc. 

Infrared luminosities are calculated for star forming clumps with assigned distances.
Figure~\ref{lbol_yso_mc} shows the distribution of $L_{IR}$ for these YSOs.
The range of observed luminosities matches the mean luminosity values for the stellar masses $<$ 25~\msun.
However, 
at the highest IR luminosities, the expectation value for model central masses is $\sim$7~\msun.

\subsection{Monte-Carlo estimates to young stellar objects masses} 
The variation of $L_{IR}$ with stellar mass 
computed in the \citet{Robitaille:2006} models and illustrated in Figure~\ref{lbol_yso_mc} enable 
a statistical estimate to the mass of the YSO object linked to an ATLASGAL clump.
For a given star with mid-infrared luminosity, $L_{IR}$, and 
luminosity error, $\sigma(L_{IR})$, a distribution of luminosities, $N(L_*)$, with an expectation value, $L_{IR}$ 
is created 
with 16384 elements 
assuming $\sigma(L_{IR})$ follows a gaussian distribution.   The largest 
source of luminosity error is due to the distance measurement rather than photometric uncertainties. 
The fractional distance errors, $\sigma_D/D$, for the ATLASGAL sources range from 0.1 to 1
\citep{Wienen:2015, Ellsworth-Bowers:2015}. 
The cumulative probability distribution, $P(M,L_{IR}|L_*)$ is constructed 
from $\zeta(M,L_{IR})$ for this set of Class I models,
\begin{equation}
P(M,L_{IR}|L_*)= \frac{\zeta(<M,L_{IR}|L_*)} {\sum\limits_{\substack{M=M_{min}}}^{M_{max}}  \zeta(M,L_{IR}|L_*)} , 
\end{equation}
where $M_{min}$=0.1~\msun\ and $M_{max}$=50~\msun\ are the minimum and maximum model YSO masses respectively. 
It describes the 
fraction of models within mass bins, $M+dM$ for $L_{IR} < L_* < L_{IR}+dL$. 
For each luminosity value in the distribution, a stellar mass, $m_*$,
 is calculated by randomly 
sampling this cumulative probability distribution. 
The assigned mass of the star, $M_*$, is the expectation value of the resultant mass distribution, 
$N(m_*)$,  and YSO mass 
uncertainties are derived from the shape of $N(m_*)$.  
The large variance of model luminosity values 
for a given central, stellar mass arises from the adopted range of disk accretion rates and 
are reflected 
in the errors in the derived stellar mass.

\begin{figure}
\centering
\includegraphics[width=\hsize]{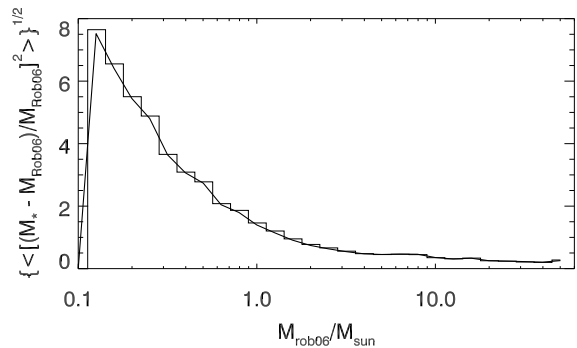}
\caption{
Variation of normalized scatter within bins of model mass for 
1\% (histogram) and 100\% (solid line)  fractional errors in the luminosity. 
The method is limited by the degeneracy of $L_{IR}$ with a range of central YSO 
masses rather than observational errors.
}
\label{model_mstar_rms}
\end{figure}

To assess the reliability of this Monte Carlo based YSO mass estimate, we apply the method to the 
Class~I model luminosities for various levels of luminosity errors.  
Figure~\ref{model_mstar_rms} shows the variation of fractional scatter, 
$(<(M_*-M_{Rob06})^2>)^{1/2}/M_{Rob06}$ with model YSO mass for both 1\% and 100\% 
fractional luminosity errors, where $M_*$ is the Monte-Carlo derived YSO mass and $M_{Rob06}$ is the model central mass. 
The congruence of these curves demonstrates that the 
method is limited by the range and diversity of the models that are 
integrated into $P(M,L_{IR}|L_*)$ rather than distance or photometric errors. 
For low masses, 
$0.1 < M_{Rob06}/M_\odot < 0.5$, the fractional 
scatter ranges from 200-750\%.  For larger masses, the fractional error narrows from 150\% at 
1~\msun\ to $\sim$20\% at 10~\msun.   While such fractional errors are too large for 
many scientific questions, these are sufficient for our more limited goals of 
defining the efficiency and star formation rate within the sample of dense clumps described 
in \S6.2. 

\begin{figure}
\centering
\includegraphics[width=\hsize]{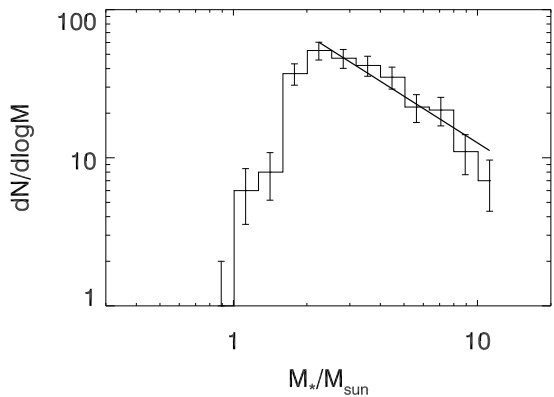}
\caption{ Mass distribution, dN/dlogM, of early stage MIPSGAL YSOs within equal sized 
logarithmic bins.  Error bars reflect counting statistics.  The solid line 
shows the error-weighted fit to the distribution for masses greater than 2~\msun\ with 
index, $\alpha_M=1.05 \pm 0.14$.
}
\label{dndm}
\end{figure}

The method is applied to the set of derived infrared luminosities 
of 290 early-stage MIPSGAL YSOs linked to ATLASGAL sources with established distances.
Table~\ref{table2} tabulates the derived mass and uncertainties for this sample.  
Also listed are the linked, ATLASGAL source name and infrared luminosity.
Figure~\ref{dndm} shows the mass distribution, dN/dlogM, for this set of YSOs.
Masses range from 1 to 10 \msun.  An error-weighted fit to the expression 
$dN/dlogM \propto M^{-\alpha_M}$ for 
masses greater than 2 \msun\ produces an index, $\alpha_M=1.05 \pm 0.14$.  The mass 
distribution is shallower than the 
the high mass IMF functional forms ($\alpha_M \sim1.3$) of \citet{Salpeter:1955} and \citet{Kroupa:2001}.
For the more luminous objects, source confusion of multiple objects within the aperture overpopulates the 
high mass bins and decreases the 
number of objects within the lower mass bins -- leading to a shallower slope in the stellar mass distribution.
Furthermore, the detected protostar is possibly
the brightest and most massive object within a very young cluster developing within the clump.  
Since the developing cluster is likely in its early stages, it is unlikely that the IMF has been 
fully sampled and the current most massive YSO could be superceded by a yet to be formed protostar. 
In any case, the distribution of the maximum protostellar mass, derived from a sample 
of Atlasgal sources with a broad range of clump masses, would not necessarily adhere to the established 
form of the IMF. 

Based on the YSO models of \citet{Robitaille:2006} and our application of the confusion 
flux limit, it is evident that our list of early-stage newborn stars samples 
a lower mass population of protostars than those compiled in previous studies. 
Both \citet{Contreras:2013} and \citet{Urquhart:2014b} used 
low sensitivity MSX IR photometry or 
 established
signatures of massive star formation 
to link newborn stars to ATLASGAL clumps. 
\citet{Csengeri:2014} used a combination of MSX and WISE photometry but without 
the application of a source completeness limit 
so their sample is biased towards brighter objects and thus, more massive sources or clusters.   
They adopt a YSO mass range of 
8-10 \msun. 
MIPSGAL photometry is more sensitive than WISE to fainter YSOs within regions of low background.
The smaller sample of ATLASGAL clumps selected by the local 24\micron\ background,  
enables linking fainter 
YSOs with lower mass while excluding bright YSOs owing to local saturation.

The total stellar mass within a clump, $M_{*,T}$ is the sum of 
masses of all linked YSOs.  For most clumps, only one Class~I YSO is identified. 
Since faint, low mass or highly obscured protostars may also be 
present within the clump, $M_{*,T}$ is 
a lower limit to the amount of YSO mass.  
An upper limit to the total mass of YSOs within the clump 
can be derived by integrating the 2-component 
initial mass function (IMF) described by \citet{Kroupa:2001}. 
\begin{equation}
M_{*,imf}=N_0\int_{0.1}^{0.5} m^{-\alpha_0+1}dm + N_1 \int_{0.5}^{M_*} m^{-\alpha_1+1} dm , 
\end{equation}
where $\alpha_0=1.3$, $\alpha_1$ = 2.3, and $M_*$ is 
the mass of the largest embedded MIPSGAL YSO within the clump. 
The constants, $N_0, N_1$, are set by $M_*$ and assumes 
continuity at the intersection of the two power laws 
such 
that $N_1=(M_*)^{\alpha_1}$ and $N_0=N_1 (0.5)^{\alpha_0-\alpha_1}=2N_1$ stars per mass interval. 
Since the 
time scale for Class~I protostars is short and the clump masses are 
small ($<$10$^4$ \msun), it is unlikely that the IMF is fully sampled so $M_{*,imf}$ 
is necessarily an upper limit to the amount of protostellar mass. 
The ratio, $M_{*,imf}/M_{*,T}$ is listed in Table~3.  The derived values can be quite large for $M_* >$ 4~\msun\ 
but are typically 6-12 for most clumps in our sample.  

\section{ATLASGAL clump properties}
The set 
of clump
properties is compiled 
for the 
sub-sample of 219 star forming ATLASGAL sources with measured distances and are fully 
resolved 
by the 19\arcsec\ beam of the 
APEX telescope.  
Clump masses are estimated from the expression
\begin{equation}
M_{cl}=\frac{S_{870} R D^2}{B_{870}(T_D)\kappa_{870}} ,
\end{equation}
where $S_{870}$ is the measured integrated flux density at 870\micron, 
$B_{870}(T_D)$ is the Planck function evaluated at this wavelength for dust temperature, 
$T_D$, assumed to be 20~K, $\kappa_{870}$ is the dust opacity set to 1.85 $cm^2/gm$, $R$
is the gas to dust mass ratio (assumed to be 100) and $D$ is the distance to the source. 
The clump radius, $r_{cl}$, is calculated from the effective angular radius, 
$\theta_{eff}$ in radians, 
$r_{cl}=\theta_{eff}D$.  The molecular gas surface density, $\Sigma_{H2}=M_{cl}/{\pi}r_{cl}^2$. 
A mean volume number density is derived assuming a spherical distribution,
\begin{equation}
n_{cl}=\frac{3M_{cl}} {4\pi \mu m_H r_{cl}^3}  ,
\end{equation}
where $\mu$=2.8 and $m_H$ is the mass of the hydrogen atom. 
The free-fall time is $\tau_{ff}=(3\pi/32G{\mu}{m_H}n_{cl})^{1/2}$. 
Finally, the clump crossing time, $\tau_{cross}=2r_{cl}/\sigma_v$,
where $\sigma_v$ is the NH$_3$ (1,1) velocity dispersion from \citet{Wienen:2012} if 
available.   
The orbital period 
of an ATLASGAL source is $\tau_{orb}=2 \pi R_G/V_\theta$, where 
$R_G$ is the Galactic radius and $V_\theta$=254 \kms\ is the azimuthal 
velocity for a flat rotation curve 
\citep{Reid:2009}.  

Errors in the time 
scales, ($\tau_{ff}, \tau_{cross}$),  are assumed to be dominated by distance errors, as these 
relate to the clump density and size respectively.
The distance uncertainty to the source dominates the error in the clump mass calculation while 
the 
assumptions of constant dust temperature and opacity can lead to 
additional errors. The 
distance errors are propagated through the calculations for $r_{cl}$, $n_{cl}$, $\tau_{ff}$, 
and $\tau_{cross}$. 
The clump properties are listed in Table~\ref{table3}. 


\section{Discussion}

\subsection{Latency of star formation in dense clumps}
The latency of star formation is the 
time interval that precedes the formation of stars within a given interstellar 
volume 
element. 
This delay could result from the required 
 development of 
gravitationally unstable, high density  
cores within a supersonic,super-Alfv\'enic medium 
or the diffusion of the magnetic field to increase local 
mass to flux ratios. 
If the volume is already in a state of large-scale gravitational collapse, then 
this latency period must be very short, $\le 2\tau_{ff}$. 

Assuming a random sampling of the star formation 
state of a clump over its lifetime, then the measured value of 
$f_*$ corresponds to the fraction of a clump lifetime, $\tau_{SF}/\tau_{cl}$,
over which newborn stars are present.  
Here, $\tau_{cl}$ is a average clump lifetime and 
$\tau_{SF}$ is the time over which a clump exhibits evidence 
for newborn stars within its domain.  
If one further assumes that star formation does not go through multiple cycles of 
 star-forming and quiescent 
stages \citep{Kruijssen:2014a},
then (1-f$_*$) corresponds to the star formation latency in units
of the clump lifetime.  
This definition for $\tau_{SF}$ is different than 
the definition 
provided by \citet{Mouschovias:2006} who define $\tau_{SF}$ as the 
time period required for the clump or molecular cloud to 
develop into a state that is susceptible for 
star formation, which is equivalent to our definition of the star formation latency. 

Our analysis of MIPSGAL YSOs and ATLASGAL clumps determines a rather small value for $f_{*}$ but 
one consistent with several previous ATLASGAL studies \citep{Contreras:2013, Csengeri:2014}.  Earlier studies 
connecting embedded protostars from the IRAS Point Source Catalog to dense core regions
identified by NH$_3$ (1,1) or (2,2) line emission, found values of f$_*$ from 33-45\% \citep{Beichman:1986, Bourke:1995}.
Less obscured T Tauri-like YSOs are linked to 1/3 of the dense cores examined by \citet{Beichman:1986} 
indicating that star formation may have 
started at an earlier stage of the core lifetime.  

We can 
not ignore the possibility that clumps without a MIPSGAL YSO may in fact
harbor faint or highly obscured YSOs with fluxes below the local, 
sensitivity limit.  Any conclusions drawn from the value of $f_*$ 
are limited to a mass range of YSOs for which our YSO list is complete. 
Based on the turnover of the YSO mass distribution shown 
in Figure~\ref{dndm} this subsample of 219 clumps is 
$\sim$90\% complete for protostars within the 
mass range 
2 $< M_*/M_\odot <$ 12.
The measured fraction of star forming clumps, corrected for star formation 
in regions of 24\micron\ saturation, is 0.31.  
Statistically, this sample of ATLASGAL clumps is engaged in star formation
 in this stellar mass range 
for only $\sim$1/3 of their lifetimes.  By complement, such clumps reside within a 
quiescent, non-star forming  
state or a state in which only low mass ($<$ 2\msun) stars are generated 
for 2/3 of their lifetimes.   
\citet{Ginsburg:2012} demonstrate that the starless phase of clumps 
forming massive star clusters 
is very short.  Such star and cluster forming clumps are likely excluded from our sample owing to the local 
24\micron\ background and non-saturation requirements.  

Recent numerical simulations of cloud formation and evolution illustrate 
clouds, clumps, and filaments in a state of hierarchical collapse 
\citep{Vazquez:2009, Ballesteros:2011}.  In this picture, the production 
of newborn stars begins 
very early within dense substructures so most or all clumps with high density contrast with 
respect to the cloud should contain 
one or more recently formed stars within several free-fall times. This leads to large values 
of $f_*$ and very short latency periods of star formation.  

These predictions would appear to be incompatible with measured low values of $f_*$
unless one or more of the following conditions apply. 
A fraction of the clumps 
may be transient features that never develop to form stars.  Adding these short-lived 
clumps to the population of clumps that do evolve to form stars violates our assumption of random 
sampling of clumps with mean lifetime, $\tau_{cl}$, and biases our measure of $f_*$ to lower values.
Furthermore, the clump lifetimes themselves could be much smaller than the lifetime 
of the overlying cloud \citep{Heyer:2015}. 
If $\tau_{SF}=2\tau_{ff}$, as implied by rapid star formation descriptions, then 
the measured value of $f_{*,c}$ implies a mean clump lifetime of $\sim$6.5$\tau_{ff}$.
Finally, if the earliest stars that 
form have low mass ($<$ 2~\msun) and therefore, are not detected 
by MIPSGAL, then $f_*$ is underestimated.  
Such a skewed YSO mass distribution in the earliest stages of a clump may be 
an essential requirement 
for the development of more massive stars \citep{Krumholz:2008}. 
The low mass stars generated in the inital stages of the clump provide a local 
heating source 
to the gas, which 
suppresses fragmentation and enables the collapse of more massive and dense sub-structures
that ultimately form high mass stars. 

\subsection{ATLASGAL clumps in the context of star formation laws} 
The data compiled in the previous sections enable an examination of the 
proposed star formation laws within the dense gas regime of the ATLASGAL clumps distributed 
throughout the Milky Way.
Previous studies have considered star forming regions within 1~kpc of the Sun that are limited in 
number ($\sim$20) and lie within a sector of the Milky Way where the environmental 
conditions such as molecular gas surface density and ambient UV radiation field may be very 
different than the inner Galaxy \citep{Gutermuth:2009, Heiderman:2010, Gutermuth:2011, Lada:2012, Heyer:2015}.  
The study by \citet{Wu:2005} evaluated the relationship between the far infrared luminosity 
and dense gas, as probed by HCN J=1-0 line emission for a sample of massive dense cores with infrared luminosities 
between 10$^3$ and 10$^7$ L$_\odot$, many of which contain ultracompact HII regions, and a smaller sub-sample of lower luminosity objects with molecular outflows.
They found a linear correlation between $L_{FIR}$ and $L_{HCN}$ 
for $L_{FIR} > 10^{4.5}$.
In our study, we have a more direct measure of \htwo\ column density gas distributed 
within  dense, compact configurations.  The mean densities listed in Table~3, are lower than the critical 
density required to excite the HCN J=1-0 
line but are generally higher than 10$^4$ \cc\ suggested by \citet{Lada:2012} as a threshold 
for the star formation process to commence. 
This sample is also larger in number than these previous studies and therefore, offers improved statistics. 
The primary limitations to our analysis are the inability to 
detect faint or highly obscured YSOs whose fluxes fall below our completion limit and 
the large uncertainties in YSO mass 
extracted from the spectral energy distribution models of \citet{Robitaille:2006}.

The star formation rate within a star-forming ATLASGAL clump is estimated 
from the total mass of associated early stage MIPSGAL YSOs,  
tabulated in 
Table~\ref{table2}, divided by the time scale for Class~I protostars. 
\begin{equation}
\dot M_* = \frac{M_{*,T} } { {\tau_{SF}} } 
\end{equation}
Studies of nearby clouds that 
are primarily forming low mass stars estimate the age of 
Class~I protostars as 0.5~Myr \citep{Evans:2009, Gutermuth:2009}.  
We adopt this value for all YSO masses. 
An upper limit to $\dot M_*$ is set by replacing $M_{*,T}$ with $M_{*,imf}$
in equation~7.   
The star formation rate surface density is 
\begin{equation}
\Sigma_{SFR} = \frac{\dot{M_*}} {\pi r_{cl}^2} 
\end{equation}
Uncertainties for $\Sigma_{SFR}$ are propagated from the error in the YSO masses tabulated 
in Table~\ref{table3} and distance errors. 

To make a fair comparison of derived star formation rates with the extragalactic measurements, 
it is important to place upper limits to the star formation rates for 
ATLASGAL sources that are not linked to 
a Class~I MIPSGAL source (D. Kruijssen, private communication).  
Such non-star forming structures are included in extragalactic CO or dust measurements 
of 
gas column density.
For each of the 291 resolved, non star forming clumps with an assigned distance, 
we calculate a 24\micron\ luminosity, $L_{24}$,
 based on the adopted 24\micron\ 
completeness limit flux of 72 mJy.  From this luminosity, the same Monte-Carlo method is applied 
to estimate a corresponding YSO mass and uncertainties using
the probability distributions, $P(M,L_{24}|L_*)$ similarly constructed from the \citet{Robitaille:2006} Class~I models. 
We assign the +2$\sigma$ mass as an upper limit for the YSO mass within each unlinked ATLASGAL clump.  
The corresponding 2$\sigma$ upper limits of $\Sigma_{SFR}$ for these non-star forming clumps are reported in Table~\ref{table4}. 

\begin{figure*} [htp]
\centering
\includegraphics[width=0.8\hsize]{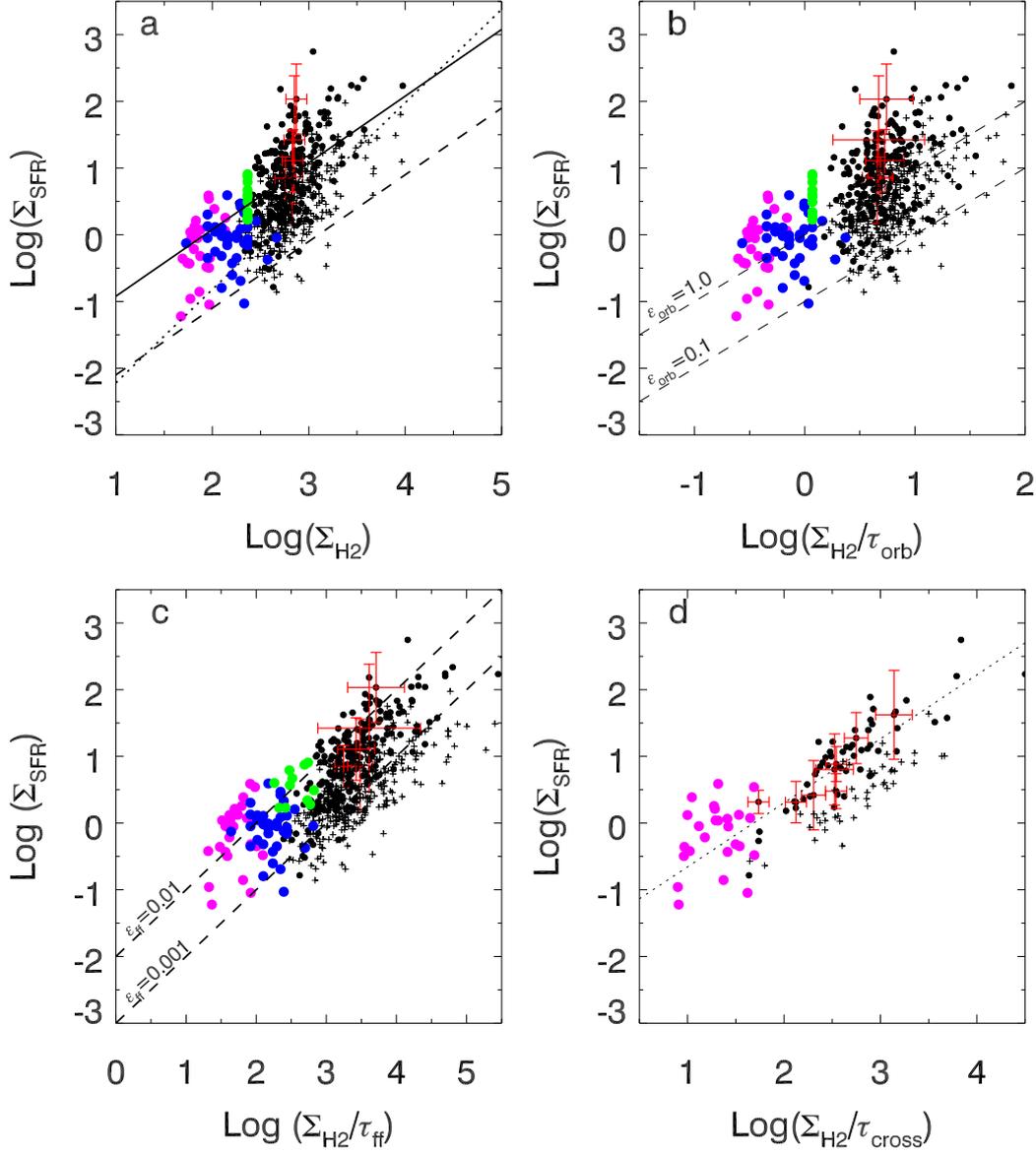}
\caption{Variation of star formation rate normalized by area, $\Sigma_{SFR}$  
with a) $\Sigma_{H2}$, b) $\Sigma_{H2}/\tau_{orb}$, 
c) $\Sigma_{H2}/\tau_{ff}$, and d) $\Sigma_{H2}/\tau_{cross}$ for resolved ATLASGAL clumps 
with measured distances. Points correspond to star forming ATLASGAL clumps (black circles and red error bars), 
2$\sigma$ upper limits to $\Sigma_{SFR}$ for non-star forming ATLASGAL clumps (black crosses) 
\citet{Lada:2012} (green), 
\citet{Evans:2014} (magenta), and \citet{Gutermuth:2009} (blue).  As discussed in the text, 
the black circles are lower limits to the true star formation rates.
In a) the lines correspond to \citet{Kennicutt:1998} (dotted), \citet{Bigiel:2008} (dashed), 
\citet{Wu:2005} (solid). In b) and c), the dashed lines correspond respectively 
to the global star formation law with  $\epsilon_{orb}$=0.1 and 1.0 and
the volumetric star formation law with  $\epsilon_{ff}$=0.001 and 0.01.
d), the dotted line 
is an error-weighted fit to the star forming ATLASGAL clumps. 
}
\label{sflaws}
\end{figure*}

The variation of $\Sigma_{SFR}$ with \htwo\ mass surface density, $\Sigma_{H2}$,  for these ATLASGAL 
sources is shown in Figure~\ref{sflaws}a. The 2$\sigma$ upper limits for the starless clumps are plotted as 
crosses.  For context, 
points for local star forming regions from \citet{Gutermuth:2009}, 
\citet{Lada:2012}, and \citet{Evans:2014} are plotted.  The star formation relationships 
determined by \citet{Kennicutt:1998} and \citet{Bigiel:2008} for galaxies are also shown. 
The detected ATLASGAL star formation rates 
lie well above the extragalactic relationships for the same value 
of $\Sigma_{H2}$, as found by \citet{Heiderman:2010} and \citet{Gutermuth:2011}. 
This 
 displacement is reduced when including the star formation rate upper limits for non star forming clumps and 
suggests that there is no large descrepancy of star formation rates for comparable gas surface density 
between resolved star forming regions and the extragalactic measurements.

No significant correlation between $\Sigma_{SFR}$ and $\Sigma_{H2}$ is present in this set of ATLASGAL clumps. 
For most of these points, there is a limited 
amount of dynamic range in $\Sigma_{H2}$ and the scatter of $\Sigma_{SFR}$ is large, which 
precludes the identification of any relationship.  
Errors in $\Sigma_{SFR}$ surely contribute to some of this scatter but variance within the clump 
population is also a source.  The 
cloud of ATLASGAL points and the data from the local star forming regions do align along 
the proposed linear relationship by \citet{Wu:2005} (solid line in Figure~\ref{sflaws}a) 
that describes the star formation 
rates in dense gas.

The global star formation law posits that the production of stars is regulated by 
large scale, radial 
processes that affect the gas over an orbital time scale \citep{Wyse:1989, Kennicutt:1998,
Tan:2000,Tan:2010, Suwannajak:2014}. 
The variation of $\Sigma_{SFR}$ with $\Sigma_{H2}/\tau_{orb}$ is shown 
in 
Figure~\ref{sflaws}b.  
Orbital periods for the solar neighborhood clouds and clumps compiled by \citet{Gutermuth:2009}, 
\citet{Lada:2012}, and \citet{Evans:2014} 
are derived for $R_G$=8.4~kpc \citep{Reid:2009}. 
There is no correlation between these quantities 
within the set of ATLASGAL points owing in part, to 
the limited range of $R_G$ values. 95\% of the star-forming clumps are located between Galactic radii 2~kpc and 8~kpc.
The cloud of points is 
aligned with those of the solar neighborhood clouds
along a linear line with $\epsilon_{orb}$$\sim$1.  This efficiency value is much larger 
than the value of $\sim$0.1 estimated by \citet{Kennicutt:1998} for disk galaxies.
While large scale processes may impact the formation of molecular 
clouds from the diffuse, atomic medium \citep{Koda:2015}, these 
do not impact the star formation rates within the dense clumps 
of molecular clouds. 

The volumetric 
star formation law considers a fixed fraction of a molecular cloud or clump, $\epsilon_{ff}$,
is converted into 
stars over a free-fall time based on the local volume density.  
Figure~\ref{sflaws}c shows the ATLASGAL star forming regions in context 
with the volumetric law.  An error-weighted fit to the ATLASGAL points shows a linear relationship 
with normalization $\epsilon_{ff}=0.003\pm0.001$, a power-law index
1.02$\pm$0.06, a reduced $\chi^2$=0.89, and a root mean square of 0.4 dex. 
The fitted value for $\epsilon_{ff}$ is $\sim$3-4 times lower than the value of 0.01 
found by \citet{Krumholz:2012} that describes star formation rates from local 
clouds to high redshift galaxies.  We emphasize that these star formation 
rates are lower limits.  Repeating the fit using the total YSO mass assuming 
a fully sampled IMF, the parameters are:
$\epsilon_{ff}=0.02\pm0.01$, a power-law index
1.14$\pm$0.06, a reduced $\chi^2$=1.2, and a root mean square of 0.5 dex. 
Since $\Sigma_{SFR,imf}$ is an upper limit, the true value of $\epsilon_{ff}$ lies somewhere 
between these two (0.003, 0.02) values.  On the other hand, inclusion of the 2$\sigma$ upper limits  for 
non-star forming clumps would bias the efficiency towards lower values.  

\citet{Elmegreen:2000} summarizes observations that suggest star formation in dense 
structures is rapid.  The production of new stars begins immediately after the structure has condensed 
from its overlying substrate and is completed within several crossing time scales. 
The dependence of the star formation rate on the crossing time scale within this set of ATLASGAL clumps is 
examined in Figure~\ref{sflaws}d.  There are 59 star forming and 44 non-star forming 
ATLASGAL clumps with velocity 
dispersions from NH$_3$ line emission that also satisfy the 24\micron\ completeness and 
resolved clump size 
requirements.  The star formation rate surface density, $\Sigma_{SFR}$ is 
well correlated with $\Sigma_{H2}/\tau_{cross}$.   An error-weighted power law fit to the 
59 points produces the following parameters:
$\epsilon_{cross}=0.03\pm0.016$, a power-law index
0.91$\pm$0.09, a reduced $\chi^2$=3.2, and a root mean square of 0.3 dex. 
While this is a lower limit to the efficiency of star forming clumps per crossing time,
inclusion of the upper limits of non-star forming clumps would lower the overall efficiency 
within dense substructures of molecular clouds.  Based on these data, star formation is well 
linked to the local crossing time of a clump but does not proceed rapidly or efficiently. 

Overall, the derived star formation rates within dense clumps traced by the thermal 
dust continuum emission favors the regulation 
of stellar production by local conditions or processes.  The strongest correlations 
are identified in Figures~\ref{sflaws}c,d in which $\Sigma_{SFR}$ depends on the dense, gas 
surface density normalized by the free-fall and clump crossing times respectively.  The low 
efficiencies of 1\% or less for $\epsilon_{ff}$ and $\epsilon_{cross}$ and 
the long latency times implied by $f_*$, 
point towards a localized process or state that slows the rate 
of star formation  or limits the amount of material that is transformed into stars 
within these dense gas configurations. 


\section{Conclusions}

We have examined the ATLASGAL and MIPSGAL source catalogs to link Class~I young stellar objects to 
their natal, massive, dense clump from which these formed.  Only 23\% of the 3494 clumps, selected 
to have low, 24\micron\ backgrounds, are found to be associated with early-stage protostars.  Accounting 
for clumps with saturated 24\micron\ emission, this fraction increases to 31\%. Using models of YSO 
circumstellar environments, we estimate the central masses of YSOs linked to ATLASGAL sources with assigned 
distances.  The YSO mass range is 1 to 10 \msun, limited on the low side by background and on the high side 
by saturation of the Mips 24\micron\ detectors.   Star formation rates are derived for ATLASGAL sources with 
distances as well as upper limits for starless clumps.  These are compared to star formation regions in the 
solar neighborhood and extragalactic relationships.  Star formation relationships that consider local gas conditions 
such as the free-fall and crossing times provide a reasonable description to this set of resolved star forming regions. 

\begin{acknowledgements}
This work was supported by the NASA
ADAP grant NNX13AF08G (MH,RG), the
Collaborative Research Center 956, funded by the \emph{Deut\-sche For\-schungs\-ge\-mein\-schaft, DFG\/}  via the SPP (priority program) (MH)
and the \emph{Deut\-sche For\-schungs\-ge\-mein\-schaft, DFG\/}  via the SPP (priority program) 
1573 'Physics of the ISM' (TCs).  The authors thank the referee for insightful comments.  We acknowledge useful discussions 
with G. Wilson, D. Kruijssen, 
and E. V\'azquez-Semadeni. 
\end{acknowledgements}

\bibliographystyle{aa} 
\bibliography{cite}

\begin{table*}
\centering
\caption{Model based protostellar mass and errors}
\label{table2}
\begin{tabular}{llcccccc} 
\hline\hline             
MIPSGAL Name & ATLASGAL Name & $L_{IR}$ & 
   $M_*$     & $M_*(-2\sigma)$ & $M_*(-1\sigma)$ & $M_*(+1\sigma)$ & $M_*(+2\sigma)$ \\
             &               & ($L_\odot$)  &
       (\msun) & (\msun) & (\msun) & (\msun) & (\msun)  \\
\hline
MG005.0424-00.0976&AGAL005.041-00.097&2.78e+00& 2.39& 0.20& 0.50& 3.98&10.00\\
MG005.2041-00.0362&AGAL005.202-00.036&4.04e+01& 5.65& 1.26& 2.51& 7.94&15.85\\
MG005.3946+00.1939&AGAL005.397+00.194&8.12e-01& 1.85& 0.13& 0.25& 3.98& 7.94\\
MG005.6181-00.0821&AGAL005.617-00.082&8.58e+01& 7.10& 2.00& 3.98&10.00&15.85\\
MG007.1654+00.1314&AGAL007.166+00.131&2.09e+01& 4.59& 1.26& 2.00& 6.31&12.59\\
MG007.3349-00.5665&AGAL007.333-00.567&1.02e+01& 3.77& 0.63& 1.26& 6.31&12.59\\
MG007.6351-00.1922&AGAL007.636-00.192&8.35e+01& 7.08& 2.51& 3.98&10.00&15.85\\
MG008.2032+00.1917&AGAL008.206+00.191&1.80e+00& 2.20& 0.16& 0.40& 3.98&10.00\\
MG008.5486-00.3394&AGAL008.544-00.341&5.95e+00& 3.06& 0.40& 1.00& 5.01&12.59\\
MG008.7038-00.4121&AGAL008.706-00.414&6.71e+00& 3.45& 0.50& 1.00& 6.31&12.59\\
MG008.7081-00.4162&AGAL008.706-00.414&1.41e+01& 4.06& 0.79& 1.58& 6.31&12.59\\
MG008.8033-00.3234&AGAL008.804-00.327&2.51e+00& 2.38& 0.20& 0.50& 5.01&10.00\\
MG008.9392-00.5303&AGAL008.954-00.532&2.01e+01& 4.59& 1.26& 2.00& 6.31&12.59\\
MG008.9555-00.5352&AGAL008.954-00.532&1.45e+01& 4.03& 0.79& 1.58& 6.31&12.59\\
MG008.9565-00.5426&AGAL008.954-00.532&1.72e+01& 4.58& 1.26& 2.00& 6.31&12.59\\
MG009.2826-00.1506&AGAL009.284-00.147&8.99e-01& 1.90& 0.13& 0.32& 3.98& 7.94\\
MG009.8474-00.0321&AGAL009.851-00.031&1.33e+01& 4.01& 0.50& 1.26& 6.31&12.59\\
MG009.8572-00.0405&AGAL009.851-00.031&1.50e-01& 1.11& 0.10& 0.16& 2.00& 5.01\\
MG009.9667-00.0206&AGAL009.966-00.021&4.71e+01& 6.15& 2.00& 3.16&10.00&15.85\\
MG010.4082-00.2013&AGAL010.404-00.201&3.01e+00& 2.46& 0.20& 0.50& 5.01&10.00\\
\hline
\end{tabular}
\end{table*}

\clearpage
\begin{sidewaystable}
\centering
\caption{Properties of resolved star forming ATLASGAL clumps with distance measure}
\label{table3}
\begin{tabular}{lcccccccccccccc} 
\hline\hline             
ATLASGAL Name & D\tablefootmark{a} & $\sigma(D)$ & Ref. & M(\htwo)  & $\sigma(M(\htwo))$ & $\Sigma_{H2}$ & $\sigma(\Sigma_{H2})$ & 
   $\Sigma_{SFR}$\tablefootmark{b} & $\sigma(\Sigma_{SFR})$ & $M_{imf}/M_{*,T}$ &  $n_{cl}$ & $\tau_{ff}$ & $\tau_{orb}$ & $\tau_{cross}$\tablefootmark{c} \\
             & \multicolumn{2}{l}\hfil(kpc)\hfil & & \multicolumn{2}{l}\hfil(\msun)\hfil & \multicolumn{2}{l}\hfil($M_{\odot}pc^{-2}$)\hfil & 
             \multicolumn{2}{l}\hfil($M_{\odot}pc^{-2}Myr^{-1}$)\hfil & &
             (10$^4$ cm$^{-3}$) & (Myr) & (Myr) & (Myr) \\
\hline
AGAL005.041-00.097 &  2.74 &  1.35 & 1 &    129 &    128 &   1037 &    259 &     38 &     55 &      8 &   5.72 & 0.13 &    133 & 0.00 \\
AGAL005.202-00.036 &  2.68 &  1.34 & 1 &     78 &     77 &    744 &    186 &    108 &    131 &     32 &   4.49 & 0.15 &    135 & 0.00 \\
AGAL005.397+00.194 &  2.62 &  1.31 & 1 &    142 &    142 &    311 &     78 &      8 &     14 &      5 &   0.90 & 0.33 &    136 & 1.23 \\
AGAL005.617-00.082 & 25.15 & 12.07 & 1 &  37030 &  35543 &    429 &    107 &      0 &      0 &     45 &   0.09 & 1.03 &    395 & 9.90 \\
AGAL007.166+00.131 & 10.25 &  0.14 & 1 &   2223 &     60 &   1274 &    319 &      5 &      4 &     23 &   1.88 & 0.22 &     51 & 1.60 \\
AGAL007.333-00.567 &  3.32 &  0.75 & 1 &    567 &    256 &    457 &    114 &      6 &      6 &     17 &   0.80 & 0.34 &    120 & 2.10 \\
AGAL007.636-00.192 &  9.29 &  0.08 & 1 &   1276 &     21 &    692 &    173 &      8 &      5 &     44 &   0.99 & 0.31 &     35 & 1.99 \\
AGAL008.206+00.191 &  2.46 &  1.43 & 2 &     90 &    105 &    622 &    155 &     30 &     50 &      7 &   3.18 & 0.17 &    139 & 0.76 \\
AGAL008.544-00.341 &  4.30 &  0.98 & 2 &    374 &    171 &    759 &    190 &     12 &     13 &     12 &   2.10 & 0.21 &     97 & 1.20 \\
AGAL008.706-00.414 & 11.68 &  0.48 & 2 &  15356 &   1267 &    486 &    122 &      0 &      0 &     10 &   0.17 & 0.75 &     86 & 0.00 \\
AGAL008.804-00.327 &  3.98 &  0.76 & 2 &    315 &    120 &    263 &     66 &      4 &      6 &      8 &   0.47 & 0.45 &    105 & 0.00 \\
AGAL008.954-00.532 & 12.84 &  0.48 & 1 &   9264 &    695 &    261 &     65 &      1 &      0 &      8 &   0.09 & 1.05 &    126 & 4.70 \\
AGAL009.284-00.147 &  4.22 &  0.68 & 2 &    683 &    221 &    401 &    100 &      2 &      3 &      6 &   0.60 & 0.40 &     99 & 0.00 \\
AGAL009.851-00.031 &  2.16 &  1.44 & 2 &    104 &    140 &    626 &    156 &     61 &    106 &     15 &   2.98 & 0.18 &    146 & 0.00 \\
AGAL009.966-00.021 & 11.61 &  0.27 & 1 &   2463 &    113 &   3054 &    764 &     15 &     12 &     36 &   6.62 & 0.12 &     86 & 0.00 \\
AGAL010.404-00.201 &  2.02 &  1.44 & 2 &    309 &    441 &    640 &    160 &     10 &     20 &      9 &   1.79 & 0.23 &    150 & 1.21 \\
AGAL010.662-00.156 &  3.20 &  1.00 & 2 &    164 &    102 &   1788 &    447 &     29 &     40 &      3 &  11.50 & 0.09 &    123 & 0.00 \\
AGAL010.742-00.126 &  3.49 &  0.50 & 1 &    781 &    225 &    470 &    117 &      3 &      3 &      7 &   0.71 & 0.37 &    118 & 2.32 \\
AGAL010.991-00.082 &  2.92 &  0.97 & 2 &    702 &    464 &    300 &     75 &      2 &      2 &      6 &   0.38 & 0.50 &    129 & 2.26 \\
AGAL011.004-00.071 &  3.45 &  0.50 & 1 &    470 &    137 &    858 &    215 &     15 &     13 &     19 &   2.26 & 0.21 &    119 & 1.09 \\
\hline
\end{tabular}
\tablebib{(1)~\citet{Wienen:2015}; (2)~\citet{Ellsworth-Bowers:2015} }
\tablefoot{
\tablefoottext{a}{\citet{Reid:2009} Galactic rotation parameters.}\\
\tablefoottext{b}{Values of $\Sigma_{SFR} <$ 0.5 $M_{\odot}pc^{-2}Myr^{-1}$ are rounded to 0. }\\
\tablefoottext{c}{$\tau_{cross}$ value equal to 0 implies that there are no available velocity dispersion measurements for 
this ATLASGAL clump.}
}
\end{sidewaystable}

\clearpage
\begin{sidewaystable}
\centering
\caption{Properties of resolved, non-star forming ATLASGAL clumps with distance measure}
\label{table4}
\begin{tabular}{lccc ccccccccc} 
\hline\hline             
ATLASGAL Name  & D\tablefootmark{a}  & $\sigma(D)$ & Ref. & M(\htwo)  & $\sigma(M(\htwo))$ & $\Sigma_{H2}$ & $\sigma(\Sigma_{H2})$ & 
   $\Sigma_{SFR}$\tablefootmark{b} & $n_{cl}$ & $\tau_{ff}$ & $\tau_{orb}$ & $\tau_{cross}$\tablefootmark{c} \\
             & \multicolumn{2}{l} \hfil(kpc)\hfil  & & \multicolumn{2}{l}\hfil(\msun)\hfil & \multicolumn{2}{l}\hfil($M_{\odot}pc^{-2}$)\hfil & 
             ($M_{\odot}pc^{-2}Myr^{-1}$) &      
             (10$^4$ cm$^{-3}$) & (Myr) & (Myr) & (Myr)  \\
\hline
AGAL005.001+00.086 &  2.75 &  1.36 & 1 &    127 &    126 &    787 &    197 &      6 &   3.81 &  0.16 &    133 & 0.00  \\
AGAL005.387+00.187 &  2.62 &  1.32 & 1 &    121 &    122 &   1217 &    304 &     10 &   7.50 &  0.11 &    136 & 0.28  \\
AGAL005.491-00.441 &  2.91 &  1.16 & 1 &    245 &    195 &    270 &     68 &      1 &   0.55 &  0.41 &    129 & 0.00  \\
AGAL005.852-00.239 &  2.80 &  1.14 & 1 &    127 &    103 &    455 &    114 &      4 &   1.67 &  0.24 &    132 & 0.00  \\
AGAL006.498-00.322 &  3.27 &  0.86 & 1 &    589 &    309 &    489 &    122 &      3 &   0.87 &  0.33 &    121 & 0.00  \\
AGAL006.564-00.319 &  3.26 &  0.86 & 1 &    361 &    190 &   1276 &    319 &     11 &   4.67 &  0.14 &    122 & 0.57  \\
AGAL008.282+00.166 &  3.06 &  1.10 & 2 &    159 &    115 &    435 &    109 &      4 &   1.40 &  0.26 &    125 & 0.00  \\
AGAL008.691-00.401 & 11.76 &  0.54 & 2 &  17897 &   1645 &   1516 &    379 &      0 &   0.86 &  0.33 &     88 & 0.00  \\
AGAL009.038-00.521 &  4.48 &  0.70 & 2 &    916 &    285 &    504 &    126 &      1 &   0.73 &  0.36 &     94 & 2.42  \\
AGAL009.796-00.707 &  3.53 &  0.54 & 1 &    364 &    111 &    688 &    172 &      2 &   1.84 &  0.23 &    117 & 0.00  \\
AGAL009.951-00.366 & 13.78 &  1.32 & 2 &   8423 &   1609 &    464 &    116 &      0 &   0.21 &  0.67 &    135 & 0.00  \\
AGAL009.981-00.386 & 13.86 &  1.38 & 2 &   5747 &   1146 &   1583 &    396 &      1 &   1.62 &  0.24 &    137 & 0.00  \\
AGAL009.999-00.034 & 11.62 &  0.27 & 1 &   1513 &     69 &   1875 &    469 &      6 &   4.06 &  0.15 &     86 & 1.43  \\
AGAL010.004-00.354 & 13.86 &  1.23 & 2 &   3646 &    648 &   1004 &    251 &      1 &   1.03 &  0.30 &    137 & 0.00  \\
AGAL010.079-00.196 &  1.95 &  0.88 & 1 &     71 &     64 &    525 &    131 &      7 &   2.78 &  0.18 &    119 & 0.00  \\
AGAL010.574-00.577 &  1.19 &  1.03 & 1 &     16 &     28 &   2384 &    596 &     95 &  56.73 &  0.04 &    207 & 0.00  \\
AGAL010.659+00.079 &  2.48 &  1.43 & 2 &    232 &    268 &    533 &    133 &      2 &   1.57 &  0.25 &    139 & 0.00  \\
AGAL010.752-00.197 &  3.24 &  0.98 & 2 &    673 &    409 &    571 &    143 &      1 &   1.02 &  0.30 &    122 & 0.00  \\
AGAL010.972-00.094 &  3.58 &  0.72 & 2 &    639 &    259 &    750 &    187 &      2 &   1.58 &  0.24 &    115 & 1.36  \\
AGAL011.001-00.372 &  1.12 &  0.98 & 1 &     47 &     82 &    809 &    202 &     11 &   6.56 &  0.12 &    206 & 0.00  \\
\hline
\end{tabular}
\tablebib{(1)~\citet{Wienen:2015}; (2)~\citet{Ellsworth-Bowers:2015} }
\tablefoot{
\tablefoottext{a}{\citet{Reid:2009} Galactic rotation parameters.}\\
\tablefoottext{b}{$\Sigma_{SFR}$ are 2$\sigma$ upper limits.} \\
\tablefoottext{c}{$\tau_{cross}$ value equal to 0 implies that there are no available velocity dispersion measurements for 
this ATLASGAL clump.}
}
\end{sidewaystable}
\end{document}